\documentclass[11pt]{article}
\usepackage[totalwidth=385pt,totalheight=620pt]{geometry}
\usepackage{amsmath,amssymb,epsf,graphics,graphicx}
\usepackage{psfrag}
\newcommand\bE{{\bar E}}

\newcommand{\CH}{{\cal H}}
\newcommand\ZZ{{\mathbb Z}}
\newcommand\RR{{\mathbb R}}

\newcommand{\ri}{\right}
\newcommand{\lf}{\left}

\newcommand\blank[1]{}

\newcommand{\fract}[2]{{\textstyle\frac{#1}{#2}}}

\newcommand\eq{\begin{equation}}
\newcommand\en{\end{equation}}
\newcommand\bea{\begin{eqnarray}}
\newcommand\eea{\end{eqnarray}}
\newcommand\nn{\nonumber}
\newcommand\ba{\(\begin{array}}
\newcommand\ea{\end{array}\)}
\newcommand{\resection}[1]{\setcounter{equation}{0}\section{#1}}

\newcommand{\NN}{{\mathbb N}}

\begin{document}
\begin{titlepage}
\vskip 0.5cm
\begin{flushright}
\end{flushright}
\vskip .7cm
\begin{center}
{\Large{\bf Quasi-exact solvability, resonances and trivial monodromy
  in ordinary differential equations}}
\end{center}
\vskip 0.8cm \centerline{Patrick Dorey$^1$, Clare Dunning$^2$ and Roberto Tateo$^3$} \vskip
0.9cm \centerline{${}^1$\sl\small Dept.\ of Mathematical Sciences,
University of Durham,} \centerline{\sl\small  Durham,  DH1 3LE, United
Kingdom\,}
\vskip 0.3cm \centerline{${}^{2}$\sl\small SMSAS, University of
Kent, Canterbury, CT2 7NF, United Kingdom}
\vskip 0.3cm \centerline{${}^{3}$\sl\small Dip.\ di Fisica 
and INFN, Universit\`a di Torino,} \centerline{\sl\small Via P.\
Giuria 1, 10125 Torino, Italy}
\vskip 0.2cm \centerline{E-mails:}
\centerline{p.e.dorey@durham.ac.uk, t.c.dunning@kent.ac.uk, tateo@to.infn.it}
\vskip 1.25cm
\begin{abstract}
\noindent
A correspondence between the sextic anharmonic oscillator and a pair
of third-order ordinary differential equations is used to investigate
the phenomenon of quasi-exact solvability for eigenvalue problems
involving differential operators with order greater than two.  In
particular, links with Bender-Dunne polynomials and resonances between
independent solutions are observed for certain second-order cases, and
extended to the higher-order problems.
\end{abstract}
\bigskip

{\small


 }

\end{titlepage}
\setcounter{footnote}{0}
\def\thefootnote{\fnsymbol{footnote}}
%
\resection{Introduction}
\label{intro}
Despite the simplicity of one-dimensional quantum-mechanical systems,
full solvability is very much an exception rather than the rule. The
archetypal example  is  the harmonic oscillator, for which
exact  solvability  breaks down completely under almost any
kind of  perturbation. Nevertheless, Turbiner \cite{Turbiner:1987nw}
found the surprising fact that in some cases, corresponding to
certain multi-parameter families of
second-order differential eigenvalue problems,
there are regions of the parameter
space  for which
a finite subset of the spectrum can be found algebraically. Turbiner and
Ushveridze \cite{TurbUsh} dubbed these models quasi-exactly solvable,
or QES.
For the QES  problems of
\cite{Turbiner:1987nw}, and almost all other examples,
the differential operators
$\CH$ act invariantly in a finite dimensional subspace
${\cal P}_{n+1}$
spanned,
possibly after a gauge transformation, by the monomials
 $1 , x, \dots , x^{n}$.

It is natural to ask whether similar ideas might apply to eigenvalue
problems involving higher-order
differential operators. Turbiner has given a general classification of
differential operators of order $k$ with finite dimensional subspace
${\cal P}_{n+1}$ \cite{Turbiner:1992} for which
$n+1$ eigenfunctions have the form of a polynomial of order at most $n$.
Examples of higher-order QES operators arising in multi-mode bosonic
Hamiltonians relevant to  nonlinear
optics may be found in \cite{Dolya:2000qy,Alvarez,
  Alvarez:2004pz, Lee:2010ak,
  Lee:2010ki}.

In this paper we consider third-order problems, motivated by a link between
a specific family of third-order differential equations and the
Schr\"odinger equation for the sextic anharmonic oscillator that was
discovered and explored in \cite{Dorey:1999pv, Dorey:2001uw}.
By tuning the parameters so that the second-order
eigenproblem is at the points
at which its spectrum is QES, the corresponding isospectral third-order
eigenproblem must also be QES. It turns out that the
quasi-exact  solvability of these
third-order differential equations is rather subtle, and even
though a finite subset of eigenvalues can be found exactly,
the most natural ansatz for the
corresponding eigenfunctions turns out not to hold.
Nevertheless, we shall exhibit a number of interesting properties of
these and related second-order eigenproblems, and
show in particular that the QES eigenfunctions are the
generating functions for sets of Bender-Dunne  polynomials  whose
zeros
correspond to the QES eigenvalues.
\resection{Isospectral second and third order eigenproblems}
\label{equiv}
Turbiner showed~\cite{Turbiner:1987nw}
that the sextic potential
\eq
\CH_2(\alpha,l) \; \psi(x) \equiv \Bigl[-\frac{d^2}{dx^2}+x^{6}+\alpha x^{2}+ \frac{l(l+1)}{x^2}
\Bigr]\psi(x)=E\;\psi(x)~,
\label{sh2}
\en
with regular boundary conditions imposed on the positive real axis at $x=0$
and $x \rightarrow \infty$ by requiring for $l>-1/2$
\eq
 \psi |_{x \rightarrow 0} = x^{l+1}(1+O(x^2))~,~~~~~\psi(x)
=O(x^{-3/2-\alpha/2} \,e^{-x^4/4}) {\rm \ as \ } x\to \infty~,
\label{bcs}
\en
is quasi-exactly solvable along the lines
\eq
\alpha=\alpha_J=-(2l+1+4J)~,~~~J=1,2,3,\dots~.
\label{QES3}
\en
The  operator $\CH_2(\alpha_J,l)$ is quasi-exactly solvable because
it acts invariantly in the  finite dimensional
subspace   $\langle f_0,f_1, \dots ,f_{J{-}1} \rangle$
where
$f_n(x) = x^{l+1} \exp(-x^4/4) \, x^{2n}~.$   The
 gauge transformation  $x^{-l-1}\exp (x^4/4) \, \CH_2 \, x^{l+1}
\exp(-x^4/4)$ and variable change $x^2 =w$
transforms the Schr\"odinger equation (\ref{sh2}) into a second-order differential
equation which acts invariantly on ${\cal P}_J \equiv \langle
1,w,... w^{J-1}\rangle   $
\cite{Turbiner:1987nw}.

A convenient way to handle the quasi-exact solvability of (\ref{sh2}) is through the
Bender-Dunne polynomials  introduced in \cite{Bender:1995rh}. The idea is to  write  a  solution
to (\ref{sh2}) in  the following factorised form
\eq
\psi(x)=
e^{-x^4/4}\,
x^{l{+}1}\,
\sum_{n=0}^{\infty}
\left({\scriptstyle -}\fract{1}{4}\right)^{n}
\frac{P_n(E,\alpha,l)}{n!\,\Gamma(n{+}l{+}3/2)}\,x^{2n}\,.
\label{bdser}
\en
For (\ref{bdser}) to solve
(\ref{sh2}), the coefficients $P_n$ must satisfy the
recursion relation
\eq
P_n(E)= E P_{n-1}(E) + 16(n-1)(n-j-1)(n+l-1/2)P_{n-2}(E)\, ,~~~(n \ge 1)
\label{rec}
\en
with  $j=j(\alpha,l)=-(\alpha{+}2l{+}1)/4$ and $P_0(E)=1$.
{}From (\ref{rec}), $P_1=E$, and $P_n$ is a polynomial of degree $n$ in
$E$, known as a Bender-Dunne
polynomial.  As long as $l\neq -n{-}3/2$ for any $n\in\ZZ^+$,
(\ref{bdser}) will yield
an everywhere-convergent series solution to (\ref{sh2}).
This solution  automatically satisfies
 the boundary condition at the origin, but at general values of
$E$, it will grow
exponentially as $x\to\infty$. However,
if $\alpha$ and $l$ are such that $j(\alpha,l)=J$ is a positive integer,
the second term on the RHS of
(\ref{rec}) vanishes at $n=J{+}1$,
and
all subsequent $P_n$  factorise:
\eq
P_{n{+}J}(E,\alpha_J,l)=P_J(E,\alpha_J,l)Q_n(E,\alpha_J,l)~,
\quad (n>0, J=-(\alpha_J{+}2l{+}1)/4\in \NN)~.
\en
Thus, if $P_J(E)$ vanishes then so do all $P_{n\ge J}(E)$ and the series
(\ref{bdser}) terminates, automatically giving a normalisable solution to
(\ref{sh2}). The $J$ roots of $P_J(E)$ are the $J$ exactly-solvable
energy levels for
the model.  For $J=1$ and $J=2$, the exactly-solvable eigenvalues  are
\eq
J=1:~~ E_0=0,~~~
J=2:~~ E_\pm= \pm 2 \sqrt{2} \sqrt{3+2 l}.
\label{QES12}
\en

For all  real values of the parameter $\alpha$ and $ l >{ -}1/2$
the  sextic eigenproblems  (\ref{sh2},\ \ref{bcs})
have up to scaling  exactly the same eigenvalues as a
 family of third-order eigenproblems \cite{Dorey:1999pv,
  Dorey:2001uw}. The relevant  third-order
differential equation is
\eq
\CH_3 \; \phi(x) \equiv  \Bigl[\frac{d^3}{dx^3}+x^{3}+\frac{L}{x^3}-G \Bigl(
 \frac{1}{x^2}\frac{d}{dx}- \frac{1}{x^3} \Bigr)\Bigr]\phi(x)=  \bE \; \phi(x)
\,
\label{tsh2}
\en
where~\footnote{In equations (\ref{GG}) and (\ref{LL}) we have corrected   an overall sign typo in the corresponding
equations of \cite{Dorey:2001uw}.}
\eq
G=2- (g_0 g_1 + g_0 g_2 + g_1 g_2) \, ,~
\label{GG}
\en
\eq
L=-2-g_0 g_1 g_2+ (g_0 g_1 + g_0 g_2+ g_1 g_2)\, ,
\label{LL}
\en
\eq
g_0+g_1+g_2=3~,
\label{kappa}
\en
and the boundary conditions on the positive real axis
are
\eq
\phi|_{x \rightarrow 0} = x^{g_1}(1+O(x^3))~,~~~~~\phi(x)= O( x^{-1} \,
e^{-x^2/2}) {\rm \ as \ } x \to \infty
\label{bc3}
\en
with $g_0<g_1<g_2$. The asymptotic condition ensures the other two
 possible behaviours of the solution at infinity are ruled out.
The isospectrality of  ${\cal H}_2,\  {\cal H}_3$ with boundary
conditions (\ref{bcs},\ \ref{bc3})  respectively was
first discussed  in
\cite{Dorey:1999pv} for $l=0$ and, with the help of results
from~\cite{Suzuki:2000fc},  generalised to $l\ne 0$  in
\cite{Dorey:2001uw}.
The result is that the eigenvalues $E , \bar E$
associated to  ${\cal H}_2,\  {\cal H}_3$
satisfy
\eq
\bE=E/\kappa~,~~\kappa=4/(3\sqrt 3)~,
\label{iso}
\en
whenever the parameters $\{ \alpha,l \}$ and $\{g_0,g_1,g_2 \}$ in the two models are related as
\eq
\alpha=2(2-g_0-g_2)~,~~~l= (2 g_2 -3 -2 g_0)/6 ~ ,
\label{alphag}
\en
and
\eq
g_0=(1 - \alpha - 6l)/4~,~~~ g_1=(1 + \alpha/2)~,~~g_2=(7 - \alpha + 6l)/4~.
\label{galpha}
\en
The result was obtained by showing that the associated
spectral determinants--functions constructed to  vanish at the eigenvalues--are proportional.  By analytical continuation from $l$ to ${-}1{-}l$, the isospectrality
result also extends to  $l\le -1/2$  \cite{Dorey:1999pv,
  Dorey:2001uw}.

If $\alpha$ is now tuned to the quasi-exactly solvable points
$\alpha=-(4J+2l+1) $ of ${\cal H}_2$ for positive integer $J$, then
via the isospectrality (\ref{iso})
the third-order problem
 ${\cal H}_3$ with
\eq
g_0=1/2+J-l~,~~~g_1=1/2-2J-l~,~~~g_2=2+ J + 2 l
\label{gqes}
\en
has a hidden QES sector and $J$ eigenvalues can be found exactly.
Since the isospectrality proof
relates the
 eigenvalues and makes no conclusions about
the  eigenfunctions, we cannot immediately state that ${\cal H}_3$ at
the points (\ref{gqes}) is
itself `fully' quasi-exactly solvable, if for full quasi-exact solvability
one would insist on being able to find algebraically not only a subset
of the eigenvalues but also the corresponding eigenfunctions.
  One might expect that  ${\cal H}_3$
 acts invariantly in a finite-dimensional subspace spanned by
 functions of the form
$x^{g_1} \exp
(-x^2/2) \, x^{n}$. If that is the case,  the QES
 eigenfunctions will  take a simple factorised form  generalising
(\ref{bdser}). We shall show that this is not in general the case.

In \cite{Dorey:2006an} and  \cite{Dorey:2007wz}, the spectral link between  $\CH_2$ and $\CH_3$  was naturally
extended to the  adjoint operator $\CH_3^{\dagger}$ to~(\ref{tsh2}):
\eq
\CH_3^{\dagger} \; \chi^{\dagger}(x) \equiv \Bigl[\frac{d^3}{dx^3}-
x^{3}+\frac{L^{\dagger}}{x^3}-G^{\dagger} \Bigl(
 \frac{1}{x^2}\frac{d}{dx}- \frac{1}{x^3} \Bigr)\Bigr]\chi^{\dagger}(x)=  -\bE \; \chi^{\dagger}(x)
\,
\label{tsh3}
\en
where $
G^{\dagger}\equiv G~, \
L^{\dagger}\equiv -L.$
The appropriate  boundary conditions are specified by
\eq
\chi^{\dagger}|_{x \rightarrow 0} = x^{g_0^{\dagger}}(1+O(x^3))~,~~~~~\chi^{\dagger}|_{x \rightarrow \infty} \rightarrow 0~,~~
(g_0^{\dagger}= 2-g_0)~.
\label{as2}
\en

It is important to note that
the eigenfunctions
$\phi_n$ of $\CH_3$  and the
 eigenfunctions $\chi^{\dagger}_n$  of $\CH_3^\dagger$
with eigenvalues $\bar E_n$ and $-\bar E_n$ respectively are
substantially different functions, not
related to each other  by simple conjugation.
Provided the roots of the
indicial equation are ordered as
$g_0 < g_1 <g_2$ then for $g_1>-1/2$ the $\phi_n$'s are square
integrable  on  $\RR^+$,
\eq
\lim_{x \rightarrow 0} \chi^{\dagger}_n(x) \phi_m(x) =O( x^{2-g_0+g_1} )\rightarrow 0~,~~~\lim_{x \rightarrow \infty} \chi^{\dagger}_n(x) \phi_m(x) \rightarrow 0~,
\en
and the  sets $\{\chi^{\dagger}_n \}$ and $\{ \phi_n \}$ can  always be
normalized such that
\eq
\langle n| m \rangle  = \int_{0}^{\infty} \chi^{\dagger}_n(x) \phi_m(x)\; dx = \delta_{n,m}~.
\label{norm}
\en
The properties  described above are reminiscent  of the well-studied
properties  of  non-self-adjoint spectral problems and   biorthogonal
systems in quantum mechanics.  These systems were introduced  and
studied in the early days of quantum mechanics, and were more
recently  revisited
in the context of $\cal{PT}$-symmetric quantum mechanical
models~\cite{Bender:1998gh}.
The reader is addressed  to \cite{Curtright:2005zk} for a recent
review of this material.

\resection{Bender-Dunne polynomials and  projective  triviality}
\label{BAe}

It is natural to ask whether  a simple factorisation similar to
(\ref{bdser}) also  characterises   the exactly-solvable  energy levels
of  $\CH_3$  and $\CH_3^{\dagger}$.
To  answer this  question we first  identify   a simple  necessary
condition for the existence of  a factorisable solution to equation
(\ref{sh2}) of  the form
\eq
\psi(x)=
x^{l{+}1}\, {\cal P}_{2J-2}(x,E)e^{-x^4/4}~,
\label{bdser1}
\en
where ${\cal P}_{2J}(x,E)$ is a polynomial of order $2J$ in $x$. The
large-$x$  behaviour  of  $\psi$  should match   the   general
WKB   result
\eq
\psi(x) = O(x^{-3/2 -\alpha/2} e^{-x^4/4}) \quad \quad x\to \infty~.
\label{WKB}
\en
The ansatz (\ref{bdser1}) agrees  with (\ref{WKB})  for $\alpha=\alpha_J=-(2l+1+4J)$,
 a result which  precisely matches the set (\ref{QES3}). Let us now assume  the
 existence of a  solution to  (\ref{tsh2}) of the form
\eq
\phi(x) =  x^{g_1} {\cal P}_{2J-2}(x,E) e^{-x^2/2}~.
\label{fac3}
\en
The relevant WKB asymptotic for the third-order ODE (\ref{tsh2}) is
\eq
\phi(x)= O(x^{-1}  e^{-x^2/2} ) \quad \quad x\to \infty~.
\label{WKB4}
\en
Comparing with (\ref{fac3}) we find
$g_1=1-2J$, a very restricted set of values compared to
(\ref{QES3}).  Starting from  equation (\ref{tsh3}) leads to a similar
conclusion. It is easy to check that when $g_1=1{-}2J$ the ansatz
(\ref{fac3}) produces a single eigenfunction with eigenvalue $E=0$ for all odd
integers $g_1$ satisfying $g_0<g_1<g_2$.  However, away from these
points we conclude that the eigenfunctions
do not take the simple factorised form (\ref{fac3}).
Although other types of  wavefunction factorisation  cannot be ruled
out by this simple argument, we have checked that $\CH_3$ is not one of the
higher-order operators that appear in  Turbiner's classification
\cite{Turbiner:1992}.
The surprise is that despite this, a subset of the
eigenvalues of ${\cal H}_3$ with parameters (\ref{gqes}) can be found
algebraically as zeros of certain  polynomials.

Since we may not have identified  the QES sector of ${\cal H}_3$ without
the link to the isospectral problem ${\cal H}_2$,
it is natural to ask
whether there are
any
alternative ways  to detect the appearance of this hidden
quasi-exact
solvability in our
third-order equations or in other models.

To answer this question  it is convenient to step back
to~\cite{Dorey:2001uw} where, amongst other results, a  series of
full and partial isospectralities for $\CH_2(\alpha,l) \equiv\CH_2$ with boundary
conditions (\ref{bcs}) were observed.
Four of them are summarized by the following diagram
\bea
\CH_2(-(4J{+}2l{+}1),l)~~~~~~ &\longrightarrow& ~~\,~\CH_2(2J{-}2l{-}1,J{+}l) \nn\\[7pt]
\Big\updownarrow \qquad\qquad~~~~&&
\qquad~~\qquad\Big\updownarrow \nn\\[9pt]
\CH_2(2J{+}4l{+}2,-J{-}\fract{1}{2})~~~ &\longrightarrow&~~~
\CH_2(2J{+}4l{+}2,J{-}\fract{1}{2}) \nn
\eea
Vertical arrows correspond to  eigenproblems that have exactly the same
eigenvalues,
while
a horizontal arrow connects two  problems that have the same
eigenvalues up to
the elimination of all the QES levels present in the left hand models.
The  two problems on the bottom row  correspond to the {\em same}\/
Schr\"odinger equation, and differ only in the boundary
condition imposed at the origin. It follows from the diagram that the `regular' eigenvalue problem
for this equation, that with the behaviour
$x^{J+1/2}$ at the origin, has exactly the same spectrum as the
irregular problem with the $x^{-J+1/2}$ behaviour at the origin, with
the exception of the first $J$ eigenvalues.

For general $l$, the eigenproblem
$\CH_2(2J{+}4l{+}2,{-}J{-}\fract{1}{2})$ also does not appear
in  Turbiner's list of QES models  \cite{Turbiner:1987nw}. Nonetheless,  it is
isospectral to the QES sextic Schr\"odinger problem and   $J$ of
its eigenvalues can be found exactly.
In~\cite{Dorey:2001uw}, after noticing  an interesting symmetry in
 the recursion relation for the  Bender-Dunne polynomials, it was
(erroneously) stated that the  appearance of QES eigenvalues in
  $\CH_2(2J{+}4l{+}2,-J{-}\fract{1}{2})$   corresponds also to
a   factorisation of the eigenfunctions in the form
(\ref{bdser}). The latter
statement can be
checked using the simple consistency  criteria
introduced above. Setting
\eq
\psi(x)=
x^{-J+1/2}\, {\cal P}_{2K-2}(x,E)e^{-x^4/4}~,
\label{bdser3}
\en
at large $x$ the wavefunction behaves as   $x^{-J-3/2+2K}e^{-x^4/4} $ while the  WKB prediction
is $  x^{-5/2-J-2l}e^{-x^4/4} $. Hence (\ref{bdser3}) is a
suitable  ansatz only for
\eq
l+1/2=-K~~,~~~K=1,2,3\dots~.
\label{strong}
\en
Again, the  constraint  (\ref{strong}) is  much stronger than
(\ref{QES3}) and
we should conclude that for general values of $l$ the QES
wavefunctions
of $\CH_2(2J{+}4l{+}2,-J{-}\fract{1}{2})$
do not take a factorised form such as (\ref{bdser3}).
We have not ruled
out that the eigenfunctions can be written in terms of other
elementary  functions, a point to which we return at the end of this
section. However, the key point  is that with the
standard techniques the quasi-exact solvability of
$\CH_2(2J{+}4l{+}2,-J{-}\fract{1}{2})$ would not be evident and
we would not know a-priori how to determine the QES eigenvalues.

Returning to the question of how to detect quasi-exact solvability in
such problems, we note that
the eigenfunction $\psi(x,E,l)$ selected  by the boundary
conditions (\ref{bcs}) is one of two
solutions
to (\ref{sh2}), characterised  by their  small-$x$ behaviour
\eq
\psi(x,E,l) = O(x^{l+1}) \quad
, \quad
\psi(x,E,{-}1{-}l) = O(x^{-l}) \quad\quad x \to 0~.
\label{psifun}
\en
 Provided the roots $l $ and ${-}1{-}l$  of the indicial
equation  do not differ by an integer, then the
 $\psi$-functions (\ref{psifun}) are linearly independent. Moreover the solutions  are
automatically {\it projectively trivial} around the origin,  by which
we mean that  for
arbitrary $E$
the monodromy of $\psi(x, E,l)$ around $x = 0$ is such that
\eq
\psi(e^{i 2\pi} x,E,l) = e^{i2 \pi (l{+}1)}
\psi(x,E,l)~.
\en
The  monodromy of $\psi(x,E,{-}1{-}l)$ follows via analytical
continuation $l\to {-}1{-}l$.

When
$l=-J-\fract{1}{2}$  the roots of the indicial equation differ
by $2J$ and so for $J$ integer
there is a `resonance' between the  two
solutions $\psi(x,J) \equiv \psi(x,E,-J{-}\fract{1}{2})$ and
$\psi(x,-J)\equiv \psi(x,E,J{-}\fract{1}{2})$. This pair
is no longer a  basis of
solutions to   ${\cal
  H}_2$ and a
 linearly independent  solution $\widetilde \psi(x,E)$
to $\psi(x,E,J)$ must be constructed. In contrast  to
the regular solution $\psi(x,-J)$, the final   solution  will in general
posses
 an additional logarithmic branch point at $x=0$   and  thus the  projective
triviality
property will therefore be lost:
\eq
\widetilde{\psi}( e^{i 2 \pi} x,E ) \ne  e^{i 2 \pi(-J+1/2 )} \widetilde{\psi}(x,E)~.
\en
We now show that the eigenfunctions corresponding to the QES eigenvalues
of
${\cal H}_2(2J{+}4l{+}2, -J-\fract{1}{2})$ do not acquire logarithmic
terms and so remain projectively-trivial. We suggest that this may be
a means of detecting the hidden quasi-exact solvability of such models.
To illustrate this  we  apply the
Bender-Dunne method to $\CH_2(2J{+}4l{+}2,-J{-}\fract{1}{2})$ by
setting
\eq
\psi(x)=
e^{-x^4/4}\,
x^{-J{+}1/2}\,
\sum_{n=0}^{\infty}
\left({\scriptstyle -}\fract{1}{4}\right)^{n}
\frac{Q_n(E,J,l)}{n!} \,x^{2n}\,.
\label{bdser2}
\en
In comparison with (\ref{bdser}),  the gamma function
has been dropped from the denominator of (\ref{bdser2}) to ensure the
coefficients in the series remain finite for all $n$ and
(\ref{bdser2}) is a
-by construction-  projectively-trivial solution $\psi(x)$.
Consequently the
polynomials  $Q_n(E)$ now satisfy
\eq
(n-J) Q_n(E)=E Q_{n-1}(E) + 16(n-1)(n+l-1/2 )Q_{n-2}(E)\,.
\label{rec2}
\en
Setting $Q_0(E)=1$, the recursion relation defines $Q_n(E)$ in terms of
 $E$ and $J$ for  $n<J$ just as before.  The first difference occurs when
 $n=J$ since the  LHS of (\ref{rec2}) vanishes. The
 RHS is a
$J^{\rm th}$-order polynomial in $E$ which must be zero. Its roots
therefore determine
 the QES
eigenvalues $E_n, \  n=1,\dots J$.  However, the recursion relation (\ref{rec2}) has left
$Q_J(E)$ unspecified and provided $l$ is such that $n{+}l{-}1/2 \ne 0$
the remaining coefficients do not
factorise. In general, all   $Q_{n>J}(E)$ will be a
function of the unknown coefficient $Q_J(E)$.   For example
\eq
Q_{J+1}(E) = EQ_{J}(E) + 16 J (J+l+1/2) Q_{J-1}(E)\,.
\label{pj}
\en
Since $Q_{J-1}(E_n) \ne 0$,  from   (\ref{pj}) we see that
the choice $Q_J(E_n)=0$ does not lead to a  truncation of  the series. More
precisely, the resulting solution (\ref{bdser2}) will always have  the desired
monodromy properties but  only for a precise  value of the constant $Q_J(E_n)$  will it be
asymptotically subdominant and satisfy the boundary condition at
infinity.

Returning to the question of determining the exact eigenfunctions of
the family $\CH_2(2J{+}4l{+}2, {-}J{-}1/2)$, we
note that for all odd integers $J$ one of the QES eigenvalues is $E=0$.
In this case
the eigenproblem can  be solved in terms of a
Whittaker function
\eq
\psi(x)= \frac{2^{J/4}
  \Gamma{(\fract{3}{4}{+}\fract{J}{2}{+}\fract{l}{2}) }}{\sqrt{\pi}
    \,x^{3/2}}\,W_{{-}\fract{J}{4}{-}\fract{1}{4} {-}\fract{l}{2},\fract{J}{4}}\lf (\frac{x^4}{2}\ri )~,
\label{odd}
\en
which  exactly matches (\ref{bdser2}) when $ E=0$ for the choice
 \eq
Q_J(0)= \frac{ (-1)^{ J/2-1/2} \,2^{5J/2}\sqrt{\pi}
  \,\Gamma{\lf(\fract{J}{2}{+}\fract{1}{2}\ri)}
  \,\Gamma{\lf(\frac{J}{2}{+}\fract{3}{4}{+}\fract{l}{2}\ri)}}{\Gamma{\lf(\frac{J}{2}\ri)}
  \,\Gamma{\lf(\fract{3}{4}{+}\fract{l}{2}\ri)}}~.
\en

Noticing that for small $J$ the solution (\ref{odd}) when $l$ is an
integer  can be written
in terms of Bessel functions multiplied by polynomials in $x$,
we were motivated  to try
an  ansatz of the form
\eq
\psi(x) = x^{3/2-J} \sum_{n=0}^\infty \left (a_n(E) \, x^{2n} \,
K_{\frac{1}{4}}\!\lf(\fract{x^4}{4}\ri)
+ b_n(E) \, x^{2n} \,K_{\frac{3}{4}}\!\lf(\fract{x^4}{4}\ri) \right)
\label{psi3}
\en
where $K_{n}(x)$ are modified Bessel functions of the second kind and
  $a_n(E), b_n(E)$ depend on $E,l$.
Acting with $\CH_2(2J{+}4l{+}2 , {-}J{-}1/2)$ for integer $l$ on (\ref{psi3})
and comparing powers of $x$,
we found
for $J=1$ that $E$ must  be zero,  the series on the RHS of (\ref{psi3})
truncated at $n=1+l$
and (\ref{psi3}) reproduces (\ref{odd}) as anticipated.
Setting $J=3$, we  solved for the
coefficients
$\{a_0, a_1 , \dots, a_{3+l}  , b_0,b_1, \dots, b_{3+l} \}$ and found
one  solution with $E=0$ and two further solutions
with eigenvalues
 $E=\pm 8 \sqrt{2+l}$, exactly reproducing the
three solutions
of $P_3(E)=0$.  Repeating this process for odd $J$,
we found that  (\ref{psi3}) generates
$J$ wavefunctions and constrains the QES eigenvalues to  be solutions
of
$P_J(E)$=0.  The series on the RHS of (\ref{psi3})
truncated at $n=J+l.$

Given the ansatz (\ref{psi3}) unexpectedly generated not just the zero
eigenvalues but all of the QES
eigenvalues for $l$ integer when
$J$ is an odd integer,
we then checked if it also works for
even integers $J$.
It turns
out the
ansatz (\ref{psi3}) indeed gave two
solutions satisfying the boundary conditions (\ref{bcs})
when $J=2$ provided
$E =\pm 2 \sqrt{2} \sqrt{3+2l}$, exactly matching (\ref{QES12}).
  We find that the ansatz (\ref{psi3})
worked for all integers $J,l \in \NN^{+}$ with  the series on the RHS of
(\ref{psi3}) truncating at
$n=2(J+l)$. However, at present we are not able to
generalise this ansatz to non-integer values of $l$.

Differential operators that act invariantly  on a subspace spanned by
polynomials multiplied by  special functions of either hypergeometric, Airy or Bessel type
 have been constructed in \cite{Dolya2}.
With the variable change $x=\sqrt 2 w^{1/4}$ and the  gauge
transformation $w^{{-}9/8} \CH_2 \,w^{5/8} $,
we find  $\CH_2 ( 2J{+}l{+}2,{-}J{-}1/2)$
becomes proportional up to an additive constant to the differential
 operator $J_5^{+}$ presented in
\cite{Dolya2} for the cases when $E=0$ and $l$ is an integer. To
reproduce the solutions 
(\ref{odd}), the invariant subspace given in \cite{Dolya2} for
$J_5^{+}$ must be extended to include Bessel functions multiplied by
rational powers of $x$.
We leave further details of  these  exact wavefunctions and the
investigation of
the cases when $l$ is not an integer to future work.

\resection{Projective  triviality and third-order QES models}
\label{BAe1}
Continuing the discussion of the last section, we now show that the
hidden QES sectors of
$\CH_3$ and $\CH_3^{\dagger}$ can be detected by using the
projective-triviality test   discussed above on
   $\chi^{\dagger} \equiv \chi^{\dagger}_{(0)}$, one of the
three linearly independent solutions $\{\chi_{(i)}^{\dagger}\}$ to
(\ref{tsh3}). These solutions are characterised  by their  small-$x$ behaviour
\eq
\chi^{\dagger}_{(i)}(x,\bE) = O(x^{g^{\dagger}_i}) \quad,\quad x \to 0
\en
where $g_i^\dagger = 2- g_i (i=0,1,2)$ are the roots of the indicial
equation. These solutions
 are  projectively trivial around the origin
\eq
\chi^{\dagger}_{(i)}(e^{i2\pi} x,\bE) = e^{i2 \pi g_i}
\chi_{(i)}^\dagger(x,\bE)
\label{pt}
\en
for
$(g_i-g_j) \notin \ZZ$  with  $i \ne j$.
The QES eigenvalues of  $\CH_3, \CH_3^\dagger$  appear when
\eq
g_0=1/2+J-l~,~~~g_1=1/2-2J-l~,~~~g_2=2+ J + 2 l~,
\label{gqes2}
\en
and
\eq
g^{\dagger}_0=3/2-J+l~,~~~g^{\dagger}_1=3/2+2J+l~,~~~g^{\dagger}_2=- J - 2 l~.
\en
Hence we have
\eq
g^{\dagger}_0=g^{\dagger}_1-3 J~~~,~J=1,2,3\dots~.
\en
Furthermore, for $l>-1/2$ and $J>0$ the ordering is $g^{\dagger}_2 < g^{\dagger}_0 < g^{\dagger}_1$, and
we are    in the presence of a resonance phenomena for the solutions $\{
\chi^{\dagger}_i \}$. Again, this circumstance
usually  leads to the  loss of the projective triviality property
(\ref{pt}) due to the appearance of logarithmic contributions to the
wavefunction. The analysis of \S\ref{BAe} suggests that
imposing projective triviality on the eigenfunctions
may be a  way to identify the set of exactly known eigenvalues.

Instead of using a Bender-Dunne  like ansatz for the wavefunction we
will construct $\chi^{\dagger}$ perturbatively using  Cheng's
method~\cite{cheng:1962}. The solution  $\chi^{\dagger}$  to
(\ref{tsh3}) is also a solution  to the equation
\eq
\chi^{\dagger}(x)=x^{2-g_0} + L[(x^3-\bE) \chi^{\dagger}(x)]~,
\label{chi0}
\en
where
\eq
L(x^p)= {x^{p+3} \over \prod_k (p+1+g_k) }~,~~~~(k=0,1,2)~.
\en
The function $\chi^{\dagger}$ can be considered as the $n \rightarrow \infty$ limit of a function
$\chi^{\dagger (n)}$ constructed  from $\chi^{\dagger (0)}=x^{2-g_0}$ using  the  following
recursion relation
\eq
\chi^{\dagger (n)}(x)= \chi^{\dagger (0)}(x) + L[(x^3-\bE) \chi^{\dagger (n-1)}(x)]~.
\en
After a single iteration, we have
\eq
\chi^{\dagger (1)}= x^{2-g_0}\Big(1 -  { \bE x^{3} \over \prod_k  (3-g_0+g_k)}+ {x^{6} \over \prod_j (6-g_0+g_k)} \Big)~.
\label{chi1}
\en
Using  (\ref{chi1})  we can   study the QES problem at $g_0-g_1=3J$ when
$J=1$. As
$(g_0-g_1) \rightarrow 3$ the second term
in the parenthesis on  the RHS of (\ref{chi1}) is in general
divergent. However   if the limit
 $(g_0-g_1) \rightarrow 3$ is taken simultaneously with
 $\bE \rightarrow 0$ so that $\bE/(3-g_0+g_1) \rightarrow C$ with $C$
 finite, the final result
 is again finite and, up to this order in the perturbative  expansion,
 $\chi^{\dagger}$ remains projectively trivial.  The result
$\bE=0 \leftrightarrow P_1(E)=0 $ corresponds precisely to the only exactly-solvable  energy  level at $J=1$.
Further, the result of a  second  iteration is
\bea
\chi^{\dagger (2)}&=&x^{2-g_0}\Big( 1-{\bE x^{3} \over \prod_k (3-g_0+g_k)} + {x^{6} \over \prod_k (6-g_0+g_k) } \nn \\
&-& {\bE x^{6} \over \prod_k (3-g_0+g_k)(9-g_0+g_k)} + { \bE^2 x^{6} \over \prod_k (3-g_0+g_k)(6-g_0+g_k)} \nn \\
&-& { \bE x^{9} \over \prod_k (6-g_0+g_k)(9-g_0+g_k)} + {x^{12} \over \prod_k (6-g_0+g_k)(12-g_0+g_k) }
\Big)~.~~~~~~
\label{chi2}
\eea
At $J=1$,  the potential divergences  again disappear in the limit $\bE/(3-g_0+g_1) \rightarrow C$,
and it is easy to check  that simultaneously all the subsequent
$\chi^{\dagger (n)}$'s  remain finite.
 It is also possible to check that the solution corresponding to   $C=0$ does not lead  to the desired
 subdominant   solution   (\ref{chi0}). The exact solution for $C=0$ is
\eq
\chi^{\dagger}(x)|_{C=0} = x^{2-g_0} {}_0F_2[\fract{1}{2},2-\fract{g_0}{2}, \fract{x^6}{216}]~,~~~(J=1, \bE=0)
\en
which indeed grows exponentially  as $x^{-1} e^{x^2/2}$.  In order to find the proper wavefunction we use the asymptotics
\eq
z^{(2-g_0)/6} {}_0F_2[\fract{1}{2},2-\fract{g_0}{2}, z] \sim {\Gamma(2-\fract{g_0}{2})  \over 2 \sqrt{3 \pi}}
z^{-1/6} e^{3 z^{1/3}} \quad \quad  z\to\infty ~,
\en
and introduce the linearly independent solution
\eq
\chi_1^{\dagger}= z^{(5-g_0)/6}
{}_0F_2[\fract{3}{2},\fract{5}{2}-\fract{g_0}{2}, z]
\en
which behaves asymptotically as
\eq
\chi_1^{\dagger}
\sim {\Gamma(\fract{5}{2}-\fract{g_0}{2})  \over 4 \sqrt{3 \pi}}
z^{-1/6} e^{3 z^{1/3}} \quad \quad z\to \infty~.
\en
Thus, the asymptotically vanishing solution is
\eq
\chi^{\dagger}(x) =  x^{2-g_0}
\left({}_0F_2[\fract{1}{2},2-\fract{g_0}{2}, \fract{x^6}{216}] -{
    \Gamma(2-\fract{g_0}{2}) \over 3 \sqrt{6}
    \Gamma(\fract{5}{2}-\fract{g_0}{2})} x^{3} \, {}_0F_2[\fract{3}{2},\fract{5}{2}-\fract{g_0}{2}, \fract{x^6}{216}] \right)~.
\label{sol}
\en
The solution  (\ref{sol}) corresponds to the choice
$ C=3 \sqrt{6} \,\Gamma(2-\fract{g_0}{2})
/  \Gamma(\fract{3}{2}-\fract{g_0}{2})$.

The case $J=2$ can be treated in a similar fashion:
as  $(g_0-g_1) \rightarrow 6$ the  $x^6$ coefficient of (\ref{chi2})
diverges. This singular behaviour  can be avoided  in the double limit
\eq
\lim_{\bE \rightarrow \bE_{\pm}, (g_0-g_1)\rightarrow 6  }\left( 1 + {\bE^2 \over 3(3 - g_0 + g_1)(3 - g_0 + g_2)} \right) / (6-g_0+g_1)  \rightarrow C
\label{dublim}
\en
provided
\eq
\bE_{\pm}= \pm 3 \sqrt{3} \sqrt{-2- g_1} =\pm 3  \sqrt{3 \over 2}  \sqrt{3+2l}~.
\label{eth}
\en
The result (\ref{eth})  matches the exact  energy levels (\ref{QES12})
provided $E/\bE=\kappa= 4/(3\sqrt 3)$ (cf.  (\ref{kappa})).  Further,
the numerator of (\ref{dublim}) is simply related to the  Bender-Dunne
polynomial $P_2(E)$ (see below) and it is possible to argue that there always exists  a
value of $C$ such that the wavefunction decays exponentially  at
large $x$.  This proves, therefore, that the roots of the Bender-Dunne
polynomials $P_2(E)$ are indeed part of the spectrum of the dual pair
(\ref{tsh2}, \ref{tsh3}).

More generally, for any $\{g_i\}$ the full Cheng solution can be written in the form
\eq
\chi^{\dagger}(x)=x^{2-g_0}\lf (  \sum_{n=0}^\infty \frac{ (-1)^n
\bar P_n(\bar{E})
   \, x^{3n}}
{  \prod_{k=0}^{2} (3n-g_0+g_k)}\ri)~,
\en
where  $\bar P_n(\bar{E})$ are  degree $n$ polynomials in $\bar
E$ that satisfy
\eq
\bar P_n(\bar{E})=\bar{E}\bar P_{n-1}(\bar{E}) +
\prod_{k=0}^{2} (3(n-1)-g_0+g_k) \bar P_{n-2}(\bar{E})
\label{rec3}
\en
with $\bar P_0=1, \bar P_1=\bar E$.
Restricting $\{g_i\}$ to the QES points (\ref{gqes2}),
 the recursion relation (\ref{rec3}) matches the
Bender-Dunne recursion relation (\ref{rec}) with
$\kappa^n \bar P_n (\bar E)  = P_n (E)$.

\section{Other models and conclusions }
\label{con}
In this paper we have reported some    progress  toward a more complete  understanding of the spectral equivalence between the sextic
anharmonic oscillator (\ref{sh2}) and the dual pair of third-order
ODEs (\ref{tsh2}) and (\ref{tsh3}), and the nature of the quasi-exact
solvability of the third-order problems that is thereby induced.
Although more work will be needed to complete the picture, and more
generally to understand  the emergence of standard and
hidden  quasi-exact
solvability  in dual  pairs
$\{ \CH, \CH^{\dagger} \}$  of higher-order differential operators,
we think that the concept of   projective triviality   should
be a useful tool
in  detecting QES sectors. In addition to the
examples  discussed in  \S\ref{BAe} and
\S\ref{BAe1},
we have discovered that
hidden quasi-exact solvability
is a property shared by many other models. For example, we have
applied  the same analysis to the $n^{\rm{th}}$ order differential
equations introduced in~\cite{Dorey:2000ma}
\eq
\lf[ (-1)^{n+1}
D(g_{n-1}-(n{-}1))\,D(g_{n-2}-(n{-}2))\,\dots\,
D(g_0) +x^{nM}\ri ]\psi(x) = E\psi(x)
\label{highD}
\en
where
\eq
D(g)=\left(\frac{d}{dx}-\frac{g}{x}\right)
\quad , \quad \sum_{i=0}^{n-1} g_i=\frac{n(n-1)}{2}
\en
   and the boundary conditions are
\eq
\psi|_{x \rightarrow 0} = O(x^{g_1})~,~~~~~\psi= O( x^{(1-n)M/2} \,
e^{-x^{M+1}/(M+1)}) {\rm \ as \ } x \to \infty
\label{bc4}
\en
with $g_0<g_1<\dots<g_{n-1}$.
These directly generalise the problems $\CH_2(0,{-}g_0)$ and
$\CH_3$.
 By imposing projective triviality on the wavefunctions of the adjoint
 problems at the resonant
points
$g_0-g_i=nJ$ for $M,J\in \NN$, we found
the  Cheng solutions for the adjoint problems are
\eq
\chi^{\dagger}(x)=x^{n-1-g_0}\lf (  \sum_{m=0}^\infty \frac{ (-1)^m P_m(E)
  \, x^{mn}}
{  \prod_{j=1}^m \prod_{k=0}^{n-1} (nj-g_0+g_k)  }    \ri)
\en
where the corresponding Bender-Dunne polynomials satisfy
\eq
P_m(E)=EP_{m-1}(E) - (-1)^M \prod_{j=1}^{M}\prod_{k=0}^{n-1} (n(j{+}m{-}M{-}1)-g_0+g_k)
  P_{m-1-M}(E)~.
\en
The QES eigenvalues  are the $J$ roots of
$P_{J}(E)=0$ and,  in general, the associated wavefunctions do not have an
elementary form.  As noted in
\S\ref{BAe} for the second-order models,  the non-QES  part of the
spectrum is precisely the spectrum  of the same differential equation
subject to a boundary condition that imposes regular behaviour of the
wavefunction at the
origin.

Finally, we should reiterate  that  for all the QES models
encountered in this paper when $g_0-g_i =nJ$, the ordering
$g_0<g_1<g_2< \dots$ of the 
solutions of the indicial equations  is not fulfilled.
 Hence, the norm (\ref{norm}) of the exactly-solvable states
 $\sqrt{\langle n|n \rangle}$ is divergent and  the corresponding
 radial eigenvalue problem is always `irregular'.
Higher-order differential equations of the form (\ref{highD}) have recently
been studied in  the context of ${\cal PT}$ symmetric quantum
mechanics and its generalisations  for even $n$ in
\cite{Bender:2008uu,Bender:2012yv}.    Additional  motivation for the
further study of
higher-order eigenproblems
of the type considered in this paper comes from their relevance to
particular integrable quantum field theories, via the ODE/IM
correspondence \cite{DTc,Sb,Dorey:2000ma,Dorey:2006an,BHK}.

\medskip
\noindent{\bf Acknowledgments --}
RT thanks Miloslav Znojil for useful discussions and kind  encouragement to finish this paper.
This project was  partially supported by an INFN grant PI11,
EPSRC grant EP/G039526/1, the
 Leverhulme Trust and the Italian MIUR-PRIN contract 2009KHZKRX-007 ``Symmetries of the Universe and of
the Fundamental Interactions''.

%
%
%

%
\end{document}